\documentclass[preprint,12pt,3p,times,sort&compress]{elsarticle}

\usepackage{amssymb}
\usepackage{amsmath}
\usepackage{color}
\definecolor{mygreen}{rgb}{0,0.7,0}

\definecolor{MMblue}{rgb}{0,0,0.7}

\usepackage[colorlinks=true]{hyperref}

\journal{Physics Letters B}

\begin{document}
\hspace*{\fill} IPPP/21/31, ZU-TH 45/21

\begin{frontmatter}



\title{Next-to-leading order QCD corrections to diphoton-plus-jet production through gluon fusion at the LHC}


\author[tor]{Simon Badger}
\author[uzh]{Thomas Gehrmann}
\author[uzh]{Matteo Marcoli}
\author[dur]{Ryan Moodie}

\affiliation[tor]{organization={Dipartimento di Fisica and Arnold-Regge Center, Universit\`{a} di Torino, and INFN, Sezione~di~Torino},%Department and Organization
            addressline={Via~P.~Giuria~1}, 
            city={10124 Torino},
             country={Italy}}
\affiliation[uzh]{organization={Physik-Institut, Universit\"{a}t Z\"{u}rich},%Department and Organization
            addressline={Wintherturerstrasse 190}, 
           city={CH-8057 Z\"{u}rich},
             country={Switzerland}}
\affiliation[dur]{organization={Institute for Particle Physics Phenomenology, Department of Physics, Durham~University},%Department and Organization
            addressline={South~Road}, 
            city={Durham},
            postcode={DH1 3LE}, 
            country={UK}}

\begin{abstract}
 We compute the next-to-leading order (NLO) QCD corrections to the gluon-fusion subprocess of diphoton-plus-jet production at the LHC. We compute fully differential distributions by combining two-loop virtual corrections with one-loop real radiation using antenna subtraction to cancel infrared divergences. We observe significant corrections at NLO which demonstrate the importance of combining these corrections with the quark-induced diphoton-plus-jet channel at next-to-next-to-leading order (NNLO).
\end{abstract}


%\item First calculation of NLO QCD corrections to this process.
%\item Significant corrections on top of NNLO $pp\to \gamma\gamma j$.


\end{frontmatter}


\section{Introduction}
\label{sec:intro}

Recent breakthroughs in two-loop amplitude technology are opening up a new range
of precision two-to-three scattering problems. Diphoton-plus-jet production
has been one of the first predictions to appear at 
NNLO in QCD~\cite{Chawdhry:2021hkp}. This progress is extremely timely
given the continually improving experimental measurements of diphoton
signatures~\cite{ATLAS:2021mbt}. Predictions for pure diphoton production have
been known to NNLO accuracy for some
time~\cite{Anastasiou:2002zn,Bern:2001df,Catani:2011qz,Campbell:2016yrh,Gehrmann:2020oec}. 
A $q_T$-resummed calculation at order $\text{N}^3\text{LL}^\prime+\text{NNLO}$
was presented recently~\cite{Neumann:2021zkb}.
Steps towards N${^3}$LO are being taken with the completion of the
three-loop amplitudes~\cite{Caola:2020dfu}. Diphoton-plus-jet signatures are
of particular importance at the LHC since they form the largest background to
Higgs production at high transverse momenta. The extra jet is necessary to
ensure a non-zero transverse momenta in the diphoton system.

The recently computed NNLO corrections~\cite{Chawdhry:2021hkp} of diphoton-plus-jet production display 
a good perturbative convergence, except in regions where the loop-mediated gluon-fusion process 
(which contributes to the cross section only from NNLO onwards) is numerically sizable compared to other contributions. 
In order to capture the full effects of the QCD corrections,
it is important to include loop-induced gluon-fusion channels from at least one
order higher in the perturbative series. These corrections are the subject of this article.

High precision two-to-three scattering problems have presented an enormous
challenge to the theoretical community. The development of new techniques and
methodology have been necessary to address several major bottlenecks that have
prevented predictions at NNLO in QCD from being completed.

One important ingredient is the two-loop amplitudes for which complete sets
of helicity amplitudes have recently been
completed~\cite{Agarwal:2021grm,Chawdhry:2021mkw,Agarwal:2021vdh,Badger:2021imn}.
These new results have been achieved thanks to a complete understanding of the
special functions
basis~\cite{Chicherin:2017dob,Gehrmann:2018yef,Chicherin:2018old,Chicherin:2020oor}
and a new range of reduction tools based in finite field
arithmetic~\cite{vonManteuffel:2014ixa,Peraro:2016wsq,Peraro:2019svx}. The end
products are fully analytic formulae which can be evaluated efficiently over
the phenomenologically relevant
phase-space~\cite{Abreu:2020cwb,Chawdhry:2020for,Abreu:2021oya,Agarwal:2021grm,Chawdhry:2021mkw,Agarwal:2021vdh,Badger:2021imn}.

Combining and integrating the amplitudes into differential cross sections
requires the subtraction of infrared divergences. To achieve this in a stable
and efficient way is an extremely hard problem and many solutions have been
proposed and applied in calculations up to NNLO. Such
algorithms often scale poorly with the number of external particles and only a
handful of examples for high multiplicity processes at NNLO currently
exist~\cite{Czakon:2021mjy,Chawdhry:2019bji,Kallweit:2020gcp,Chawdhry:2021hkp}.

For the process considered in this article, the infrared divergences are only at
NLO. However, since the real radiation involves $2\to4$
one-loop squared amplitudes, the automated numerical algorithms are tested in extreme
phase-space regions. The leading order (LO) QCD contributions to the gluonic subprocess were first considered in Ref.~\cite{deFlorian:1999tp} based on the compact one-loop five-gluon amplitudes~\cite{Bern:1993mq}.

Our paper is organised as follows. We first review the computational setup,
discussing the amplitude-level ingredients and antenna subtraction method used to
cancel infrared divergences. We then present results for the NLO corrections to
differential cross sections at the 13 TeV LHC. We study the perturbative convergence in both
transverse momentum and mass variables as well angular distributions in rapidity and
the Collins-Soper angle before drawing our conclusions.

\section{Computational setup}
\label{sec:setup}

We consider the scattering process
\begin{equation}
  g g \to \gamma \gamma g + X
  \label{eq:process}
\end{equation}
at a hadron collider.  As the process is loop-induced, the LO
contribution is at order $\alpha_s^3$ and involves the integration of a one-loop amplitude squared.  
The NLO QCD corrections are computed by combining the
two-loop virtual corrections to the $2\to3$ process with the $2\to 4$ processes
with an additional unresolved parton: $gg\to \gamma \gamma gg$ and $gg\to \gamma \gamma q\bar q$. 
Pictorially, we can represent the parton level cross sections up to NLO in QCD as,
\begin{align}
  \sigma_{gg\to \gamma\gamma g + X}^{\rm NLO} =&
  \int d\Phi_3 \left|%
  \raisebox{-6mm}{\includegraphics[width=1.5cm]{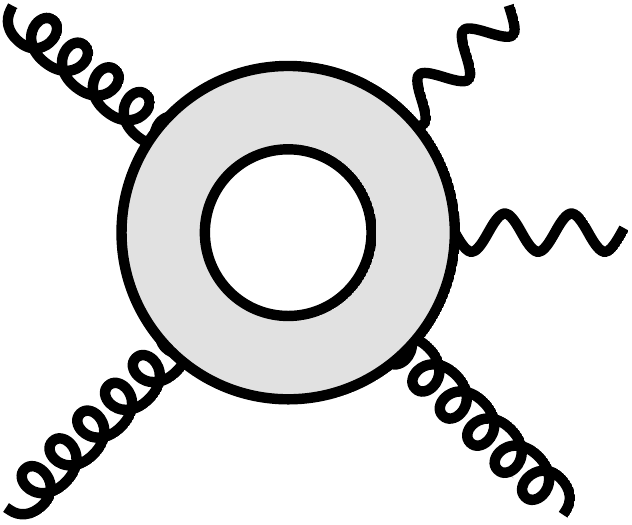}}%
  \right|^2 +
  \int d\Phi_3 2 {\rm Re}\left(%
  \raisebox{-6mm}{\includegraphics[width=1.5cm]{amppic-3g2a1L}}%
  ^\dagger\cdot%
  \raisebox{-6mm}{\includegraphics[width=1.8cm]{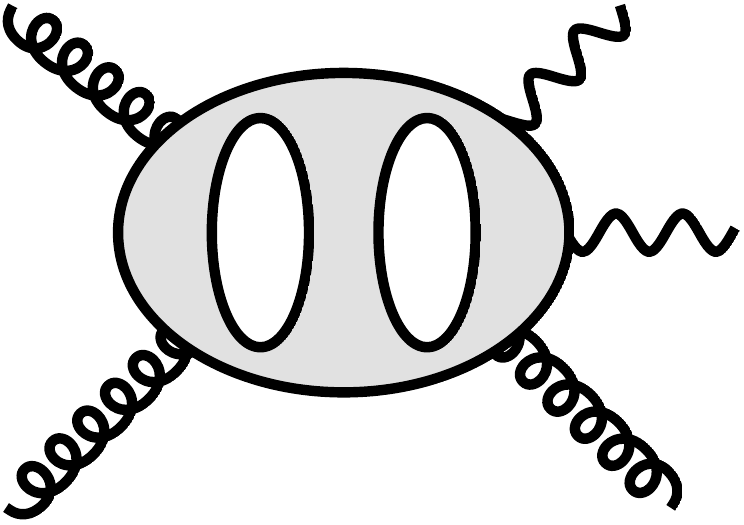}}%
  \right) + \nonumber\\&
  \int d\Phi_4 \left|%
  \raisebox{-6mm}{\includegraphics[width=1.5cm]{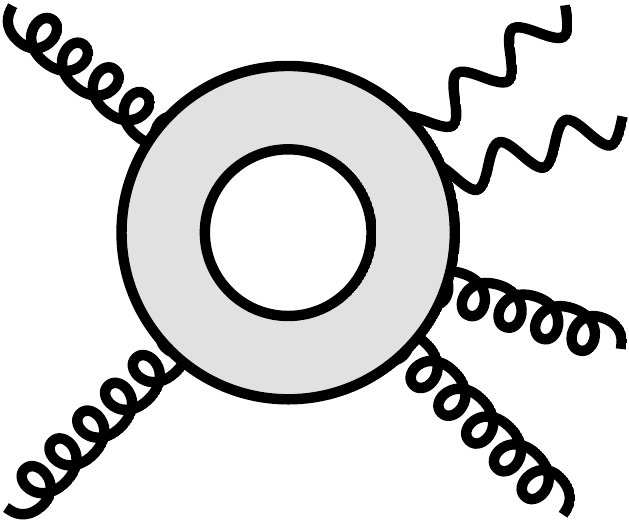}}%
  \right|^2 +
  \int d\Phi_4 \left|%
  \raisebox{-6mm}{\includegraphics[width=1.5cm]{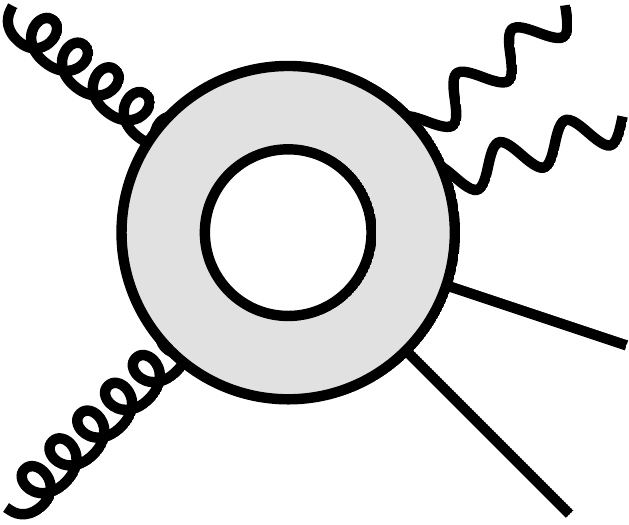}}%
  \right|^2
  + \mathcal{O}(\alpha_s^5),
  \label{eq:xsdef}
\end{align}
where $d\Phi_n$ represents the on-shell phase-space measure for $n$ massless final state particles. The one-loop amplitude for $q\bar{q}gg\gamma\gamma$ indicates the loop contribution in which the photons couple to an internal fermion loop. The observable process $pp \to \gamma\gamma j$ also includes channels where the photons couple to an external quark pair. The expansion up to the NNLO of $pp \to \gamma\gamma j$ includes terms up to $\mathcal{O}(\alpha_s^3)$ and so the contributions coming from Eq.~\eqref{eq:xsdef} are technically N${}^3$LO. However, due to the large gluon flux at high energy hadron colliders, such contributions can be significant.

The one-loop amplitudes for the LO process and the real correction are finite, since the corresponding tree-level processes vanish. The renormalised two-loop five-particle amplitude contains explicit infrared divergences generated by the integration over the loop momenta, while the one-loop six-particle amplitudes exhibit a divergent behavior when a final-state parton becomes unresolved. The divergences cancel in the final result, as established by the KLN theorem, and a finite remainder of the virtual amplitudes can be defined 
using QCD factorization~\cite{Catani:1998bh}. In our calculation, this cancellation is performed using the
antenna subtraction method~\cite{Gehrmann-DeRidder:2005btv,Daleo:2006xa,Currie:2013vh}. 
The method extracts the 
infrared singular contributions from the real radiation subprocess, and combines their integrated form with the 
virtual subprocess, thus enabling their numerical integration using Monte Carlo methods, performed here in the \textsc{NNLOjet} framework. 
The QCD structure of the process under consideration is very similar to Higgs-plus-jet production in gluon fusion, which has been computed 
previously~\cite{Chen:2014gva,Chen:2016zka} using antenna subtraction,  and identical antenna subtraction terms are applied here. 

The infrared-finite remainders of the 
two-loop amplitudes have recently been computed~\cite{Badger:2021imn} using a basis of pentagon
functions~\cite{Chicherin:2017dob,Gehrmann:2018yef,Chicherin:2020oor}, which permit efficient and reliable numerical 
evaluation~\cite{Chicherin:2020oor}. The full colour and helicity summed expressions are obtained from
the \textsc{NJet} amplitude library. Within \textsc{NJet}, a dimension scaling test is performed for each phase-space point to 
assess the accuracy of the evaluation. If the test is unsuccessful, the point is recomputed in
higher precision. We set a three digits accuracy threshold for this test, which guarantees a stable result without
significantly affecting the performance.

The one-loop six-particle amplitudes are
obtained using a combination of implementations from the \textsc{OpenLoops2}~\cite{Buccioni:2019sur}
generator and from the generalised unitarity~\cite{Bern:1994zx,Bern:1994cg,Britto:2004nc} approach within \textsc{NJet}~\cite{Badger:2012pg}. We use an improved version of \textsc{OpenLoops2} in combination with the new extension \textsc{Otter}. \textsc{Otter}~\cite{otter:2020} is a tensor integral library based on the {\em on-the-fly reduction}~\cite{Buccioni:2017yxi} of \textsc{OpenLoops2} and on stability improvements described in Ref.~\cite{Buccioni:2019sur}.
This new version of \textsc{OpenLoops2} allows for a stable computation of the needed one-loop squared amplitudes in deep infrared regions. Internally, \textsc{Otter} uses double-precision scalar integrals that are provided by \textsc{Collier}~\cite{Denner:2014gla,Denner:2016kdg}, as well as quad-precision scalar integrals provided by \textsc{OneLoop}~\cite{vanHameren:2010cp}.
Minor modifications were made in \textsc{NJet} to avoid de-symmetrisation over the two photons and allow for a pointwise correspondence with the subtraction terms.
To compute the one-loop amplitude $gggg\gamma\gamma$, the
\textsc{OpenLoops} implementation was generally more efficient, but for
exceptional phase-space points it was necessary to use the high precision (32 digits)
implementation within \textsc{NJet}. For the $q\bar{q}gg\gamma\gamma$ channel, we used
\textsc{NJet}, which allowed for a straightforward selection of the required loop contribution. We note that this amplitude is also available within \textsc{OpenLoops2} and we checked that the two implementations agree.

The amplitude-level ingredients have been validated in all relevant collinear and soft limits by checking their convergence towards the 
respective antenna subtraction terms. 

\section{Results}
\label{sec:results}

For the numerical evaluation of our NLO results on the gluon-induced diphoton-plus-jet process, we apply the same kinematical cuts  as were 
used for the NNLO calculation~\cite{Chawdhry:2021hkp} of the quark-induced processes. These represent a realistic setup relevant 
for physics studies at the 13 TeV LHC. 
The cuts are as follows:
\begin{itemize}
  \item minimum photon transverse momenta and rapidities: $p_T(\gamma_1) > 30$ GeV, $p_T(\gamma_2) > 18$ GeV and $|\eta(\gamma\gamma)|<2.4.$
  \item smooth photon isolation criterion~\cite{Frixione:1998jh}  with $\Delta R_0 = 0.4$, $E_T^{\rm max} = 10$ GeV and $\epsilon=1$.
  \item minimal invariant mass of the photon pair: $m(\gamma\gamma) \geq 90$ GeV.
  \item minimal separation of the photons: $\Delta R_{\gamma\gamma} > 0.4$.
  \item minimal transverse momentum of the photon pair: $p_{T}(\gamma\gamma) > 20$ GeV.
\end{itemize}
We consider kinematical distributions in the following diphoton variables: 
transverse momentum of the diphoton system $p_{T}(\gamma\gamma)$, pair invariant mass $m_{\gamma\gamma}$,
diphoton total rapidity $|y(\gamma\gamma)|$ and rapidity difference $\Delta y(\gamma\gamma)$, as well as Collins-Soper angle
 $\left|\phi_{CS}(\gamma\gamma)\right|$~\cite{Collins:1977iv} and azimuthal decorrelation $\Delta\phi (\gamma\gamma)$. 
For these distributions, no jet requirement is applied, as done in Ref.~\cite{Chawdhry:2021hkp}, since the 
transverse momentum cut on the diphoton system is already sufficient to avoid NNLO-like 
configurations where all final-state QCD partons become unresolved. 

Our numerical results use the NNLO set of the \textsc{NNPDF3.1} parton distribution functions~\cite{NNPDF:2017mvq} throughout, thus allowing a 
straightforward comparison with the existing NNLO results~\cite{Chawdhry:2021hkp} in the quark-initiated channels.  The strong
coupling constant is evaluated using \textsc{LHAPDF}~\cite{Buckley:2014ana}, with $\alpha_{s}\left(m_Z\right)=0.118$. The  electromagnetic 
coupling constant is set to $\alpha=1/137.035999139$. Monte Carlo integration errors are below $1\%$ on average and not displayed in the plots.

The uncertainty on 
our theory predictions is estimated by a seven-point variation of the renormalisation and 
factorisation scales around a central value, chosen in dynamical manner on event-by-event basis to 
be 
\begin{equation}
 \mu_F = \mu_R = \frac{1}{2}m_{T} = \frac{1}{2} \left( m^2(\gamma\gamma)+p_T^2(\gamma\gamma)\right)^{1/2}
\end{equation}

\begin{figure}[t]
  \begin{center}
    \includegraphics[width=0.49\columnwidth]{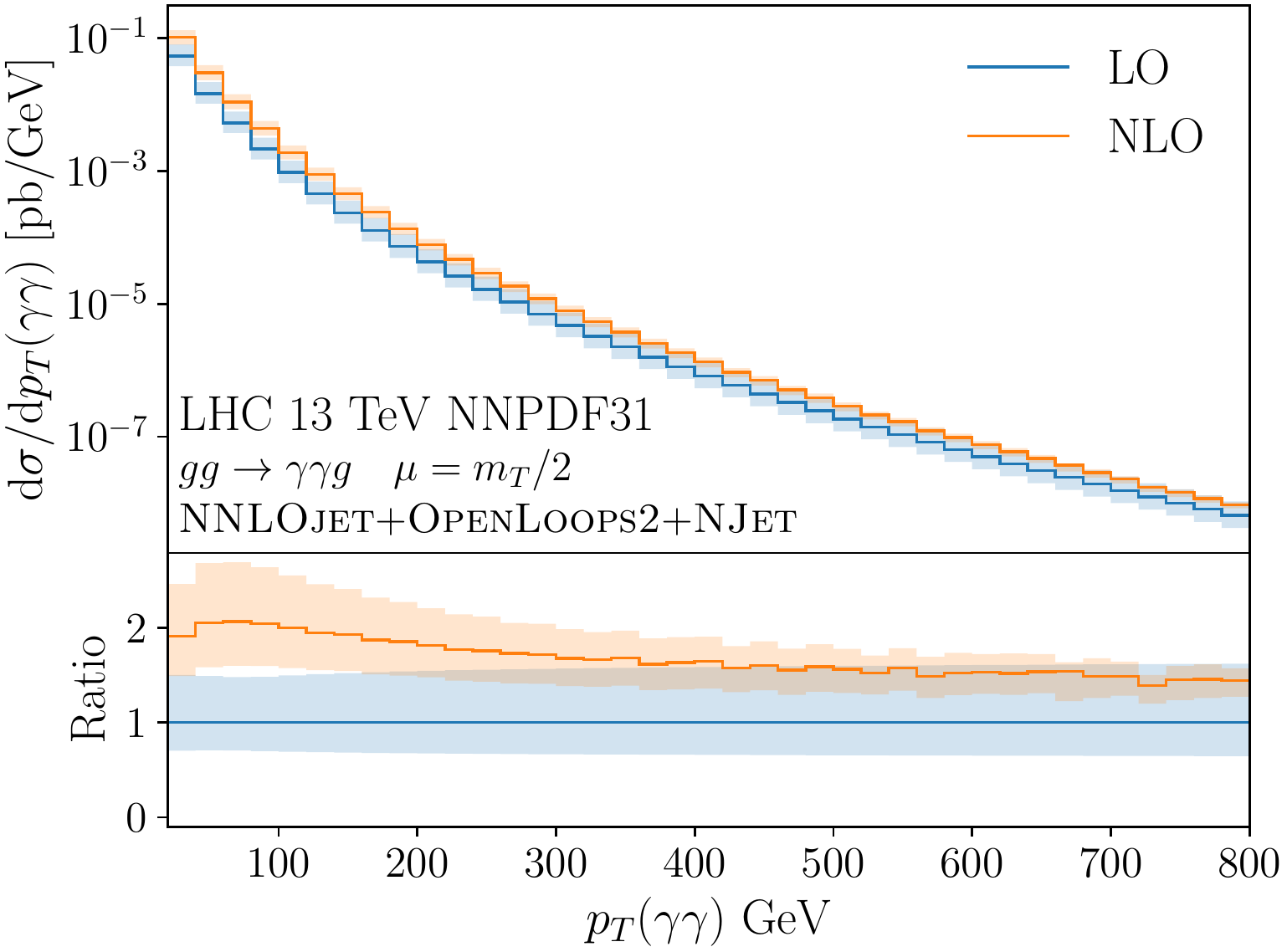}
     \includegraphics[width=0.49\columnwidth]{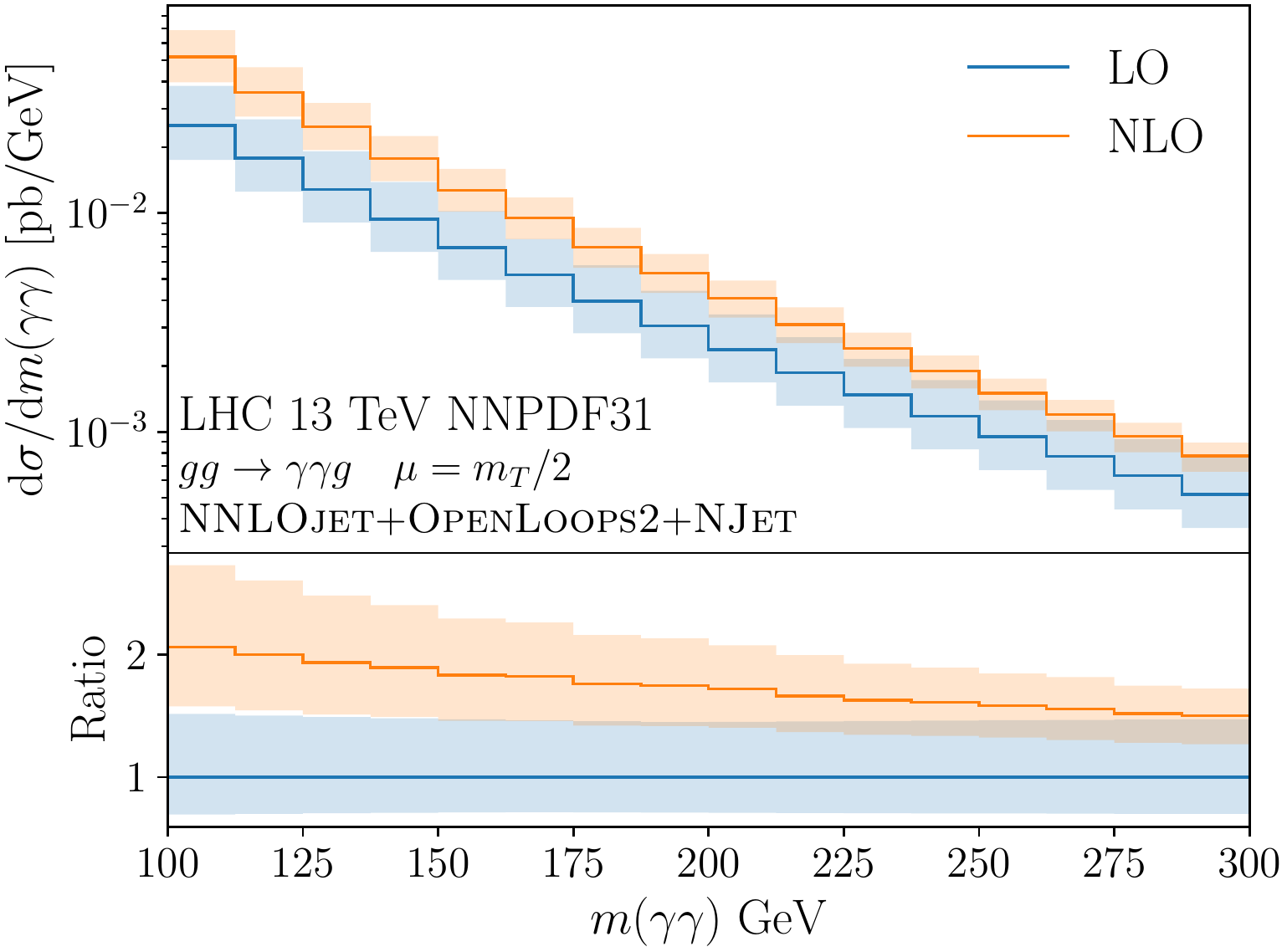}
  \end{center}
  \caption{Differential distributions in the transverse momentum $p_T(\gamma\gamma)$ (left) and invariant mass 
  $m(\gamma\gamma)$ (right) of the diphoton system.}
  \label{fig:pt_maadist}
 \begin{center}
       \includegraphics[width=0.49\columnwidth]{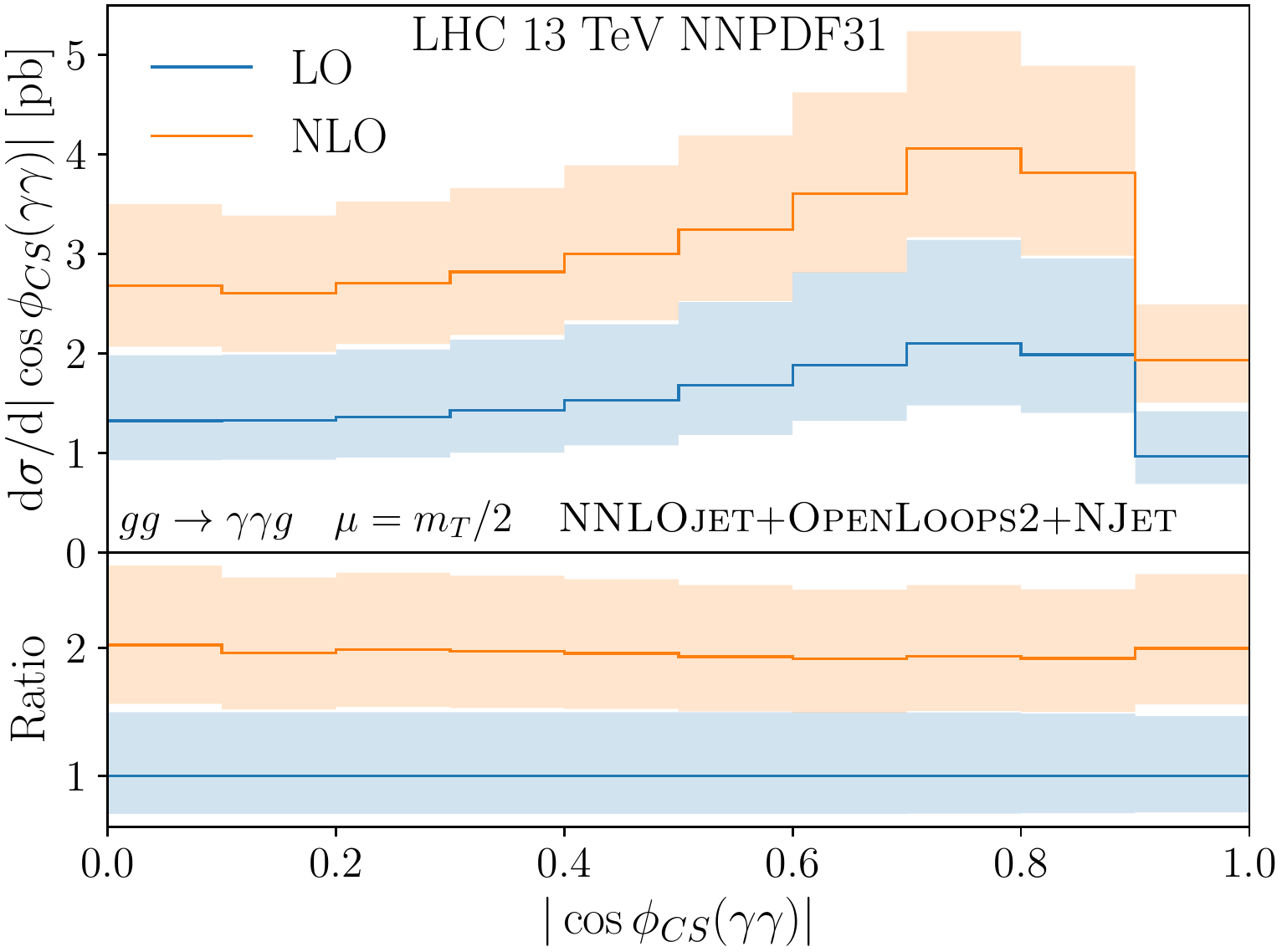}
\includegraphics[width=0.49\columnwidth]{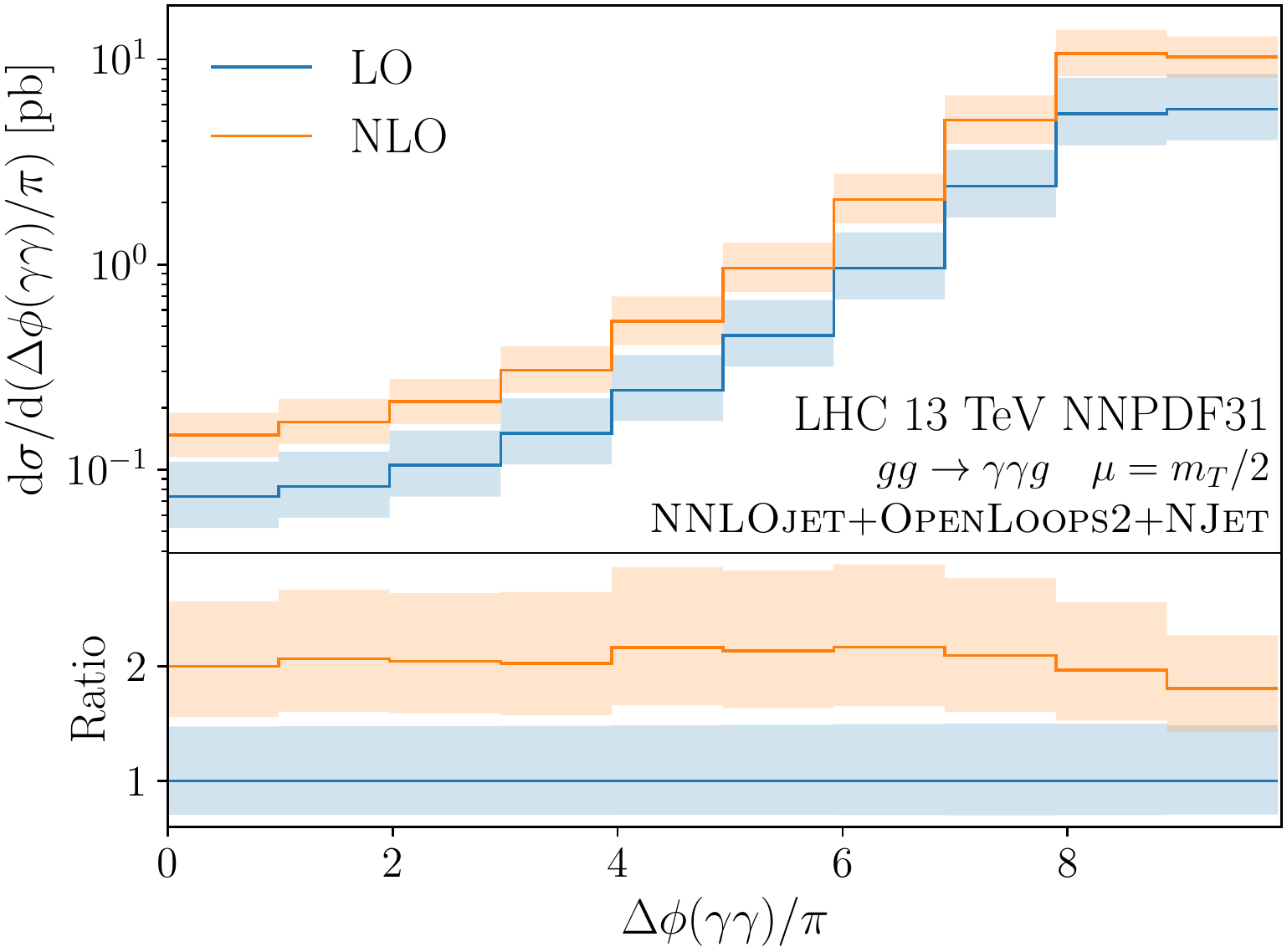}
  \end{center}
  \caption{Differential distributions in the Collins-Soper angle $\left|\cos \phi_{CS}(\gamma\gamma)\right|$ (left) the azimuthal decorrelation $\Delta \phi(\gamma\gamma)$ (right) of the diphoton system.}
  \label{fig:phidist}
\end{figure}
\begin{figure}[t]
 %\end{figure}
% \begin{figure}[h]
  \begin{center}
       \includegraphics[width=0.49\columnwidth]{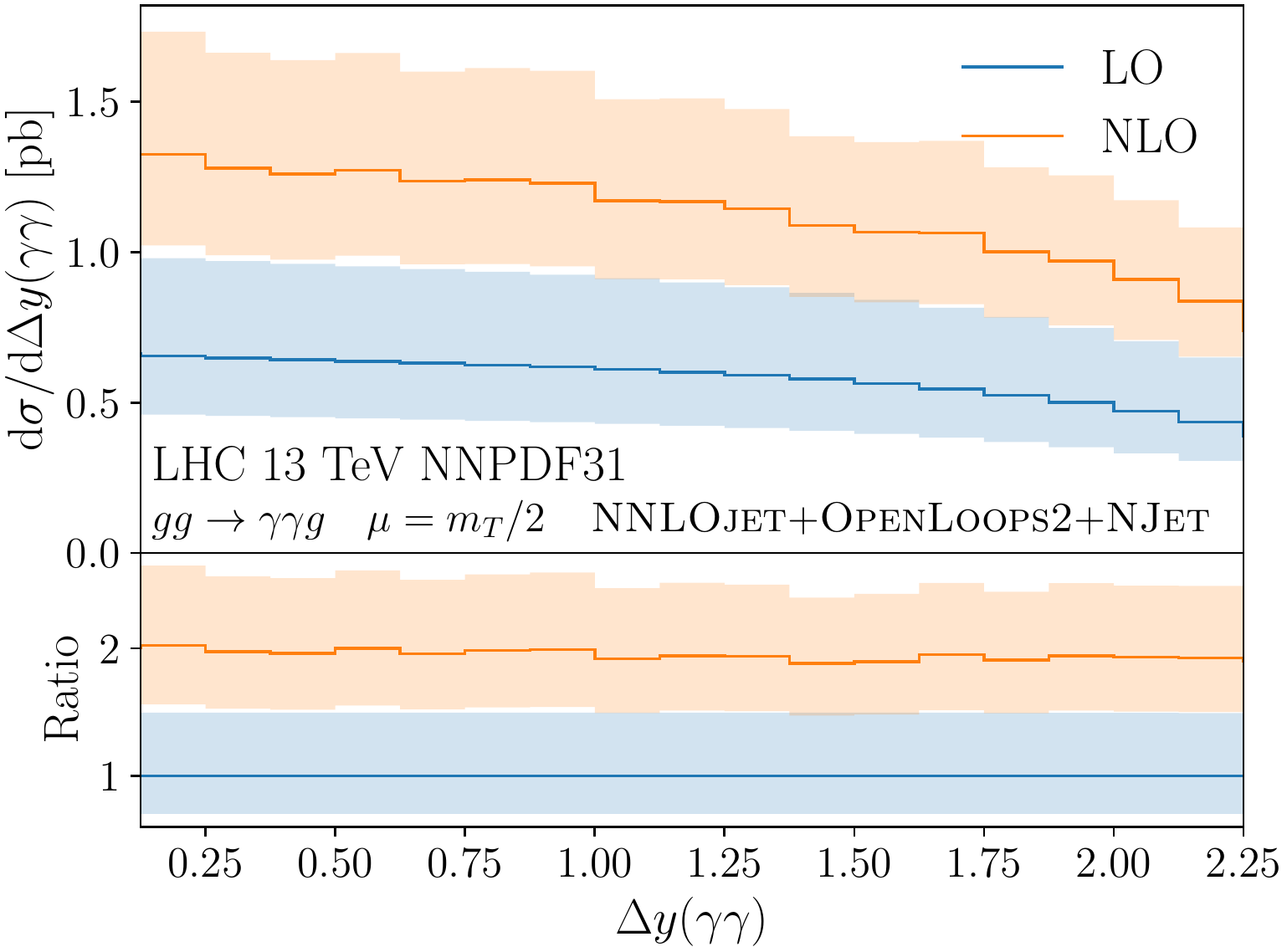}
\includegraphics[width=0.49\columnwidth]{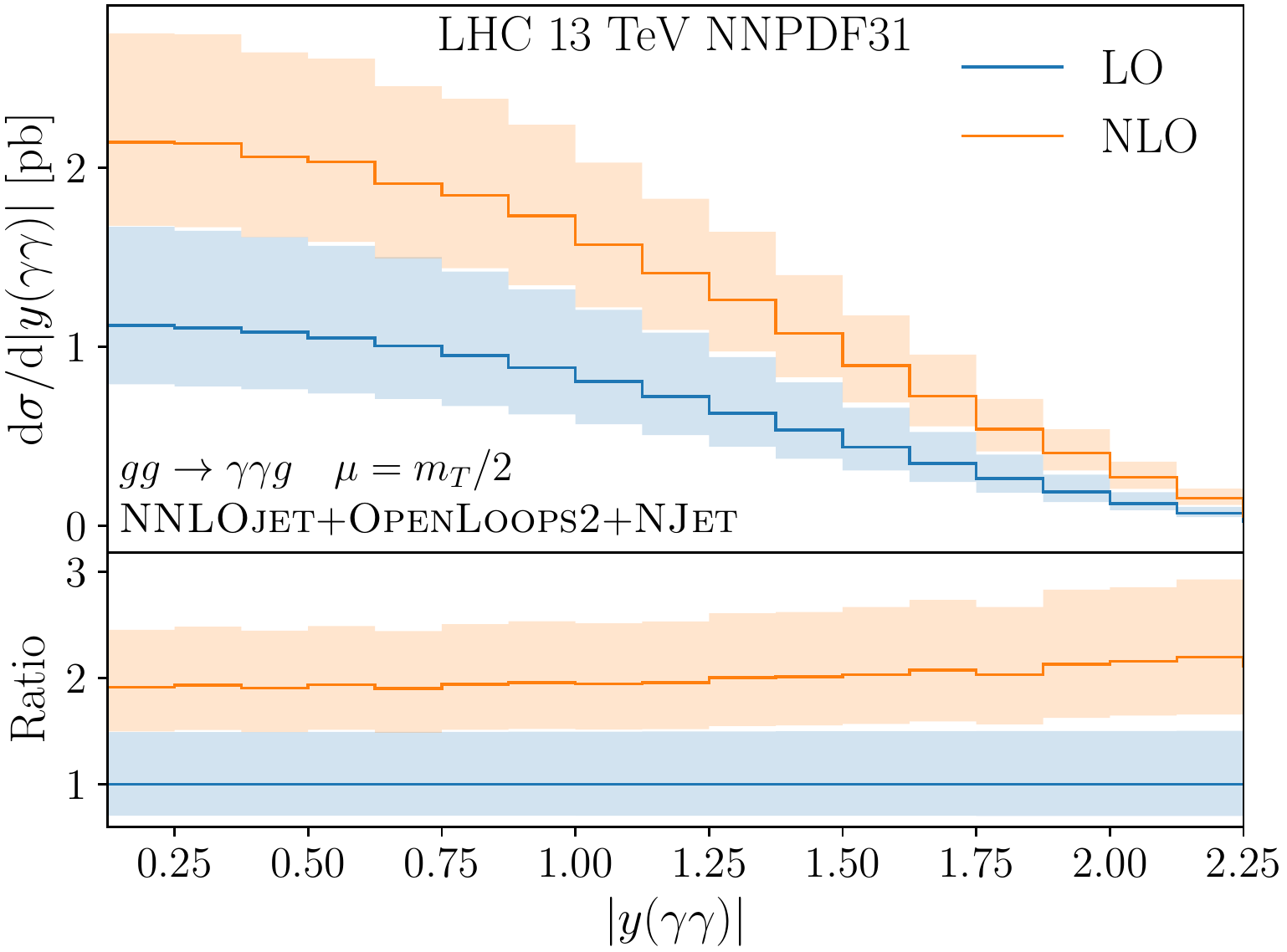}
  \end{center}
  \caption{Differential distributions in the diphoton rapidity difference $\Delta y(\gamma\gamma)$ (left) and the diphoton total 
  rapidity $|y(\gamma\gamma)|$ (right). }
  \label{fig:ydist}
\end{figure}

Figures~\ref{fig:pt_maadist}--\ref{fig:ydist} display the theory predictions for the different single-differential distributions 
in the diphoton variables. We observe the NLO corrections to be very sizable, often being comparable in size to the LO predictions. The 
corrections are largest at low $p_T(\gamma\gamma)$ or at low invariant mass, Figure~\ref{fig:pt_maadist}, where the NLO/LO ratio reaches 2
and NLO and LO uncertainties fail to overlap, while the ratio
is smoothly decreasing towards values of 1.5 for large $p_T(\gamma\gamma)$ or $m(\gamma\gamma)$, with overlapping scale uncertainty bands 
above $p_T(\gamma\gamma)=200$~GeV or $m(\gamma\gamma)=175$~GeV. 

The integrated 
cross section is dominated by the region of  low $p_T(\gamma\gamma)$ or low $m(\gamma\gamma)$, such that distributions that are differential only in geometrical 
photon variables, Figures \ref{fig:phidist}
and \ref{fig:ydist}, display typically near-uniform NLO/LO ratios of 2, and no overlap of the LO and NLO scale uncertainty bands. 
Visually, the scale uncertainty bands at NLO and LO appear to be of comparable width in all distributions. However, owing to
the large size of the NLO corrections, the relative scale uncertainty is reduced from about 50\% at LO to 30\% at NLO. 
 \begin{figure}[t]
  \begin{center}
       \includegraphics[width=0.49\columnwidth]{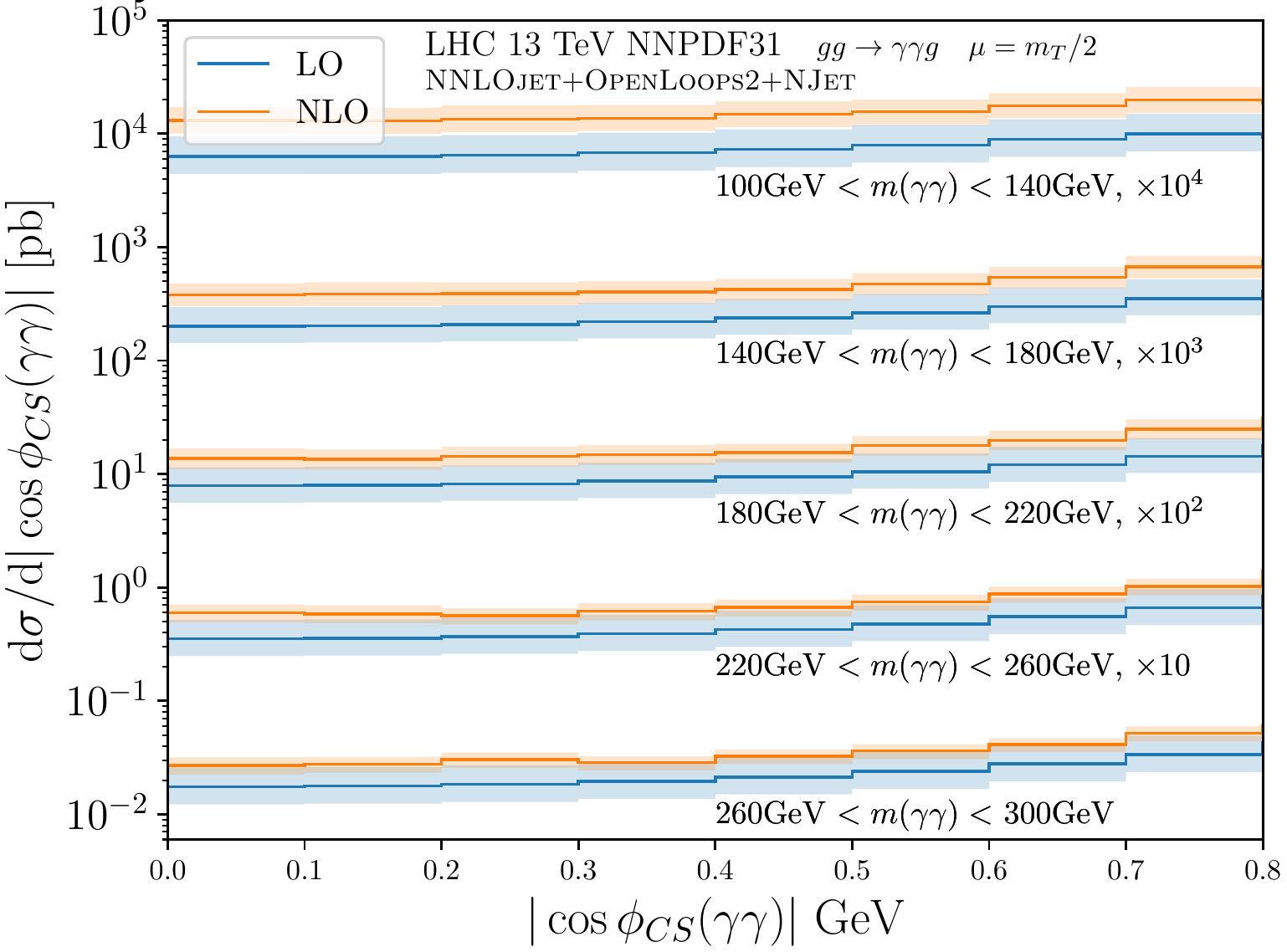}
\includegraphics[width=0.49\columnwidth]{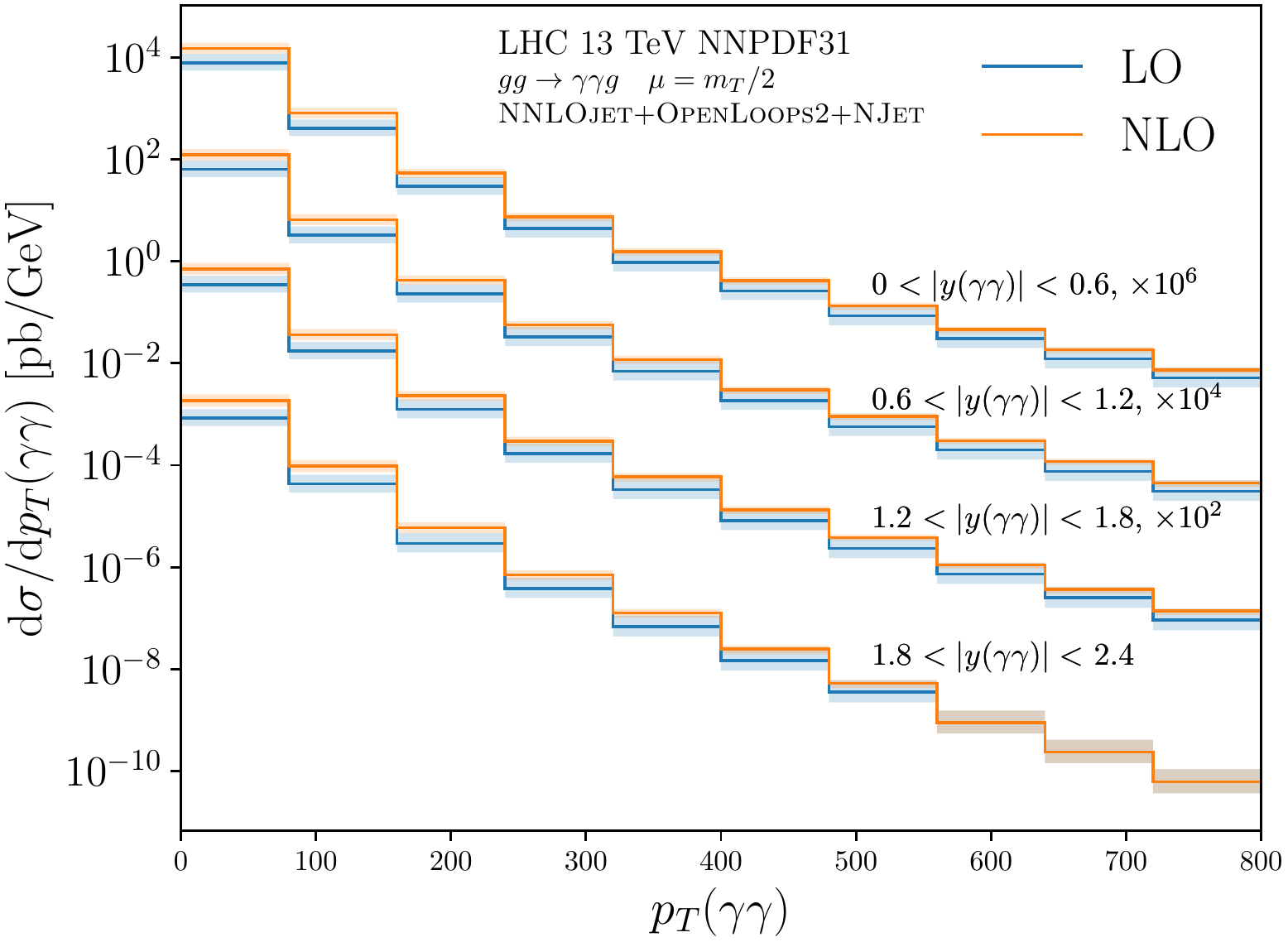}
  \end{center}
  \caption{Two-dimensional differential distributions in the diphoton invariant mass $m(\gamma\gamma)$ and 
  Collins-Soper angle $\left|\phi_{CS}(\gamma\gamma)\right|$ (left) and in the 
  diphoton rapidity $|y(\gamma\gamma)|$  and transverse momentum $p_T(\gamma\gamma)$ (right). }
  \label{fig:2Ddist}
%\end{figure}
%\begin{figure}[h]
  \begin{center}
       \includegraphics[width=0.49\columnwidth]{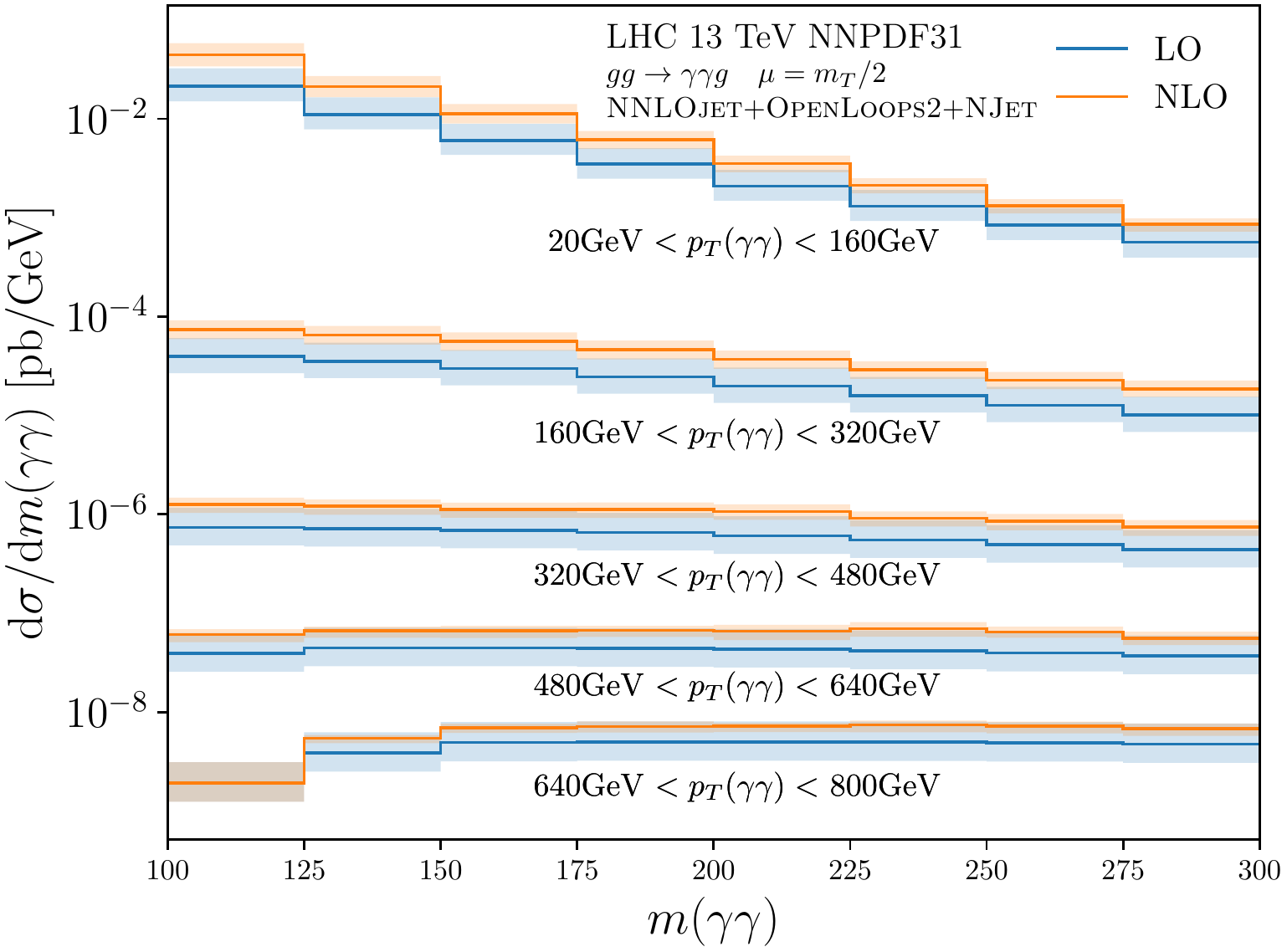}
\includegraphics[width=0.49\columnwidth]{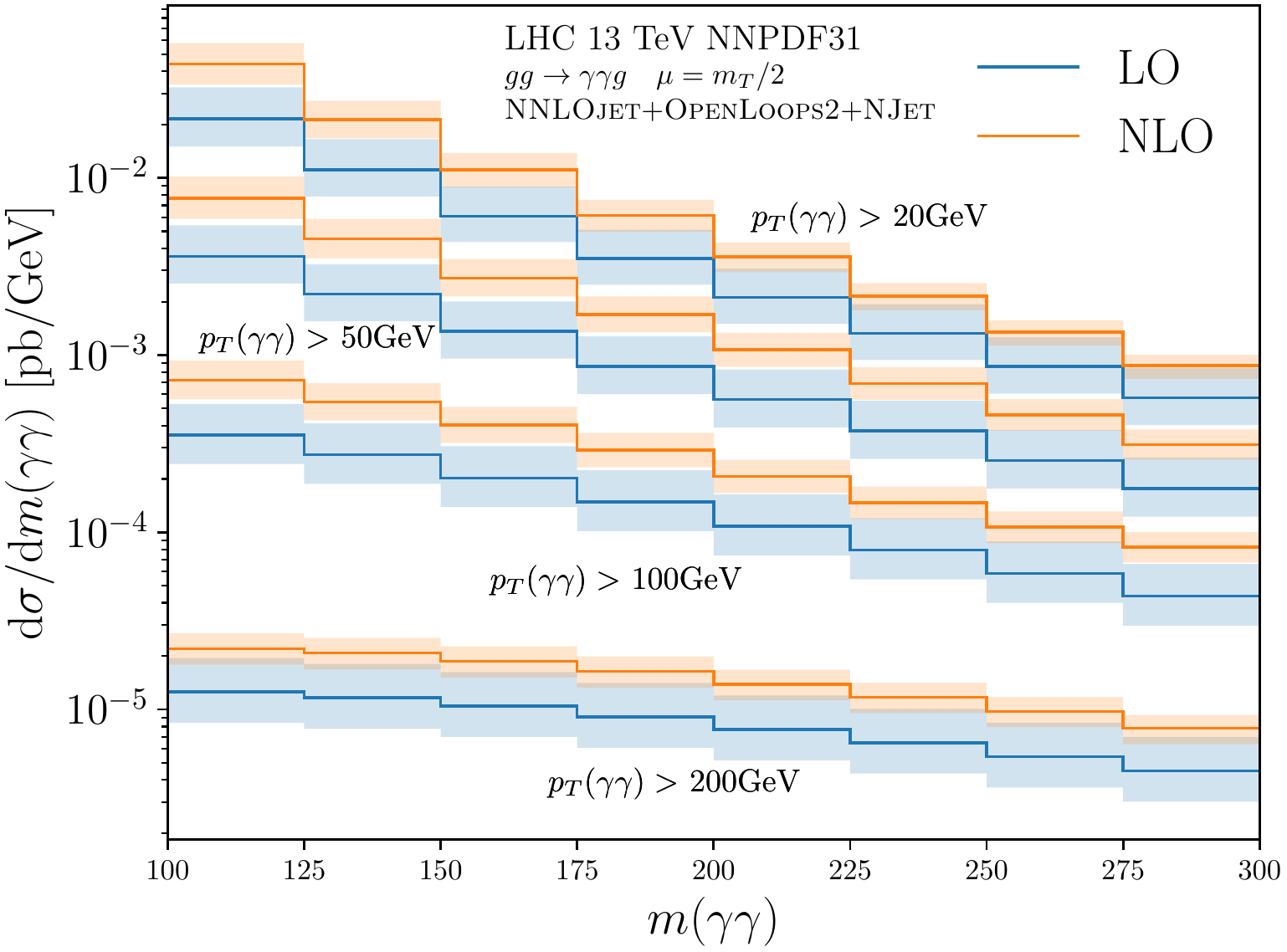}
  \end{center}
  \caption{Two-dimensional differential distributions in the diphoton transverse momentum $p_T(\gamma\gamma)$  
  and invariant mass $m(\gamma\gamma)$, in bins in $p_T(\gamma\gamma)$ (left) and for varying lower $p_T(\gamma\gamma)$-cut (right)}
  \label{fig:2Ddist2}
\end{figure}

By inspecting the two-dimensional differential distribution in $m(\gamma\gamma)$ and $\left|\phi_{CS}(\gamma\gamma)\right|$, Figure~\ref{fig:2Ddist} (left), we observe that the relative 
magnitude of the NLO 
corrections decreases with increasing $m(\gamma\gamma)$, while the corrections remain uniform in $\left|\phi_{CS}(\gamma\gamma)\right|$ for all bins in $m(\gamma\gamma)$. The 
two-dimensional differential distribution in $|y(\gamma\gamma)|$ and $p_T(\gamma\gamma)$ also shows the decrease of the corrections towards larger $p_T(\gamma\gamma)$. The decrease is more 
pronounced at forward rapidity than at central rapidity. 

Considering two-dimensional distributions in  $p_T(\gamma\gamma)$ and $m(\gamma\gamma)$, Figure~\ref{fig:2Ddist2}, largely reproduces the features of the one-dimensional 
distributions of Figure~\ref{fig:pt_maadist}, both for distributions in bins of $p_T(\gamma\gamma)$ or for varying lower cut in $p_T(\gamma\gamma)$. The only novel feature is a 
non-uniform shape in $m(\gamma\gamma)$ for the highest bin in $p_T(\gamma\gamma)$ (lowest curves in left Figure~\ref{fig:2Ddist2}), which is indicative of the onset of 
large logarithmic corrections in $\log(m(\gamma\gamma)/p_T(\gamma\gamma))$ in this range. 

The numerical size of the NLO corrections and the scale uncertainties at LO and NLO are comparable to what was observed in 
inclusive Higgs boson production in gluon fusion~\cite{Djouadi:1991tka} or in the Higgs boson transverse momentum distribution in
gluon fusion~\cite{deFlorian:1999zd,Ravindran:2002dc}. These processes are mediated through a 
heavy top quark loop and are very similar to the diphoton-plus-jet production considered here in 
terms of kinematics and initial-state parton momentum range. The pathology of the NLO corrections observed here is thus not that surprising after all;
it does however indicate the potential numerical importance of corrections beyond NLO.  

The Born-level $gg\to \gamma\gamma g$ subprocess (corresponding to the LO in our results) 
contributes to the full diphoton-plus-jet production as part of the NNLO corrections. 
Corrections to this order were computed most recently~\cite{Chawdhry:2021hkp}: these were observed to be moderate and within the 
scale uncertainty of the previously known NLO results for most of the kinematical range, where they also led to a substantial 
reduction of the scale uncertainty at NNLO. 
Only at low $p_T(\gamma\gamma)$ or low $m(\gamma\gamma)$, larger positive corrections and an increased scale uncertainty were observed~\cite{Chawdhry:2021hkp}. These 
effects could be identified to be 
entirely due to the contribution of the $gg\to \gamma\gamma g$, which only starts to contribute from NNLO onwards, and it was anticipated in
Ref.~\cite{Chawdhry:2021hkp} that NLO corrections to the  $gg\to \gamma\gamma g$ (which form a subset of the N${^3}$LO corrections to 
the full diphoton-plus-jet process) could help to stabilise the predictions in the relevant kinematical ranges. 

Our results demonstrate that this is not the case. The absolute scale uncertainty on the gluon-induced process does not decrease from LO to NLO, and the 
NLO correction is of about the size of the LO contribution. Consequently, inclusion of the NLO corrections to the $gg\to \gamma\gamma g$
into the full NNLO diphoton-plus-jet process will further enhance the predictions at low $p_T(\gamma\gamma)$ or low $m(\gamma\gamma)$, thereby further 
elongating them from the previously known order, and will leave the scale uncertainty band largely unchanged.

\section{Conclusions}
\label{sec:concl}

In this article, we have presented the NLO QCD corrections to the diphoton-plus-jet
production in the gluon-fusion channel for the first time. The loop-induced
process requires the evaluation of six-point one-loop real emission amplitudes
and full-colour five-point two-loop virtual amplitudes. To the best of our
knowledge it is the first time that five-point two-loop amplitudes with the full colour
information have been integrated to provide fully differential cross section predictions
relevant for the LHC experiments.

Using a realistic set of kinematic cuts and simulation parameters, we find
significant corrections at NLO. This is particularly relevant at low values of
$p_T(\gamma\gamma)$ and $m(\gamma\gamma)$. Since angular observables
such as rapidity and the Collins-Soper angle are inclusive over the energy
variables, one observes significant NLO corrections across the full parameter
range. Double differential distributions further highlight this feature, which
is reminiscent of the perturbative convergence observed in other gluon-induced
processes such as inclusive Higgs production and the Higgs boson transverse
momentum distribution. The relative scale uncertainty is reduced by the higher order corrections,
although in absolute terms the scale uncertainty does not decrease from LO to NLO in the low 
$p_T(\gamma\gamma)$ and $m(\gamma\gamma)$ regions.

This work demonstrates the importance of a combined prediction for quark-induced
and gluon-induced diphoton-plus-jet signatures for future precision studies at the LHC.

\section*{Acknowledgements}
We would like to thanks Johannes Henn, Xuan Chen, Alexander Huss and Simone Zoia 
for numerous discussions and interesting input in the course of this project. Help with \textsc{OpenLoops2} 
from Jean-Nicolas Lang and Federico Buccioni is gratefully acknowledged.
This research has been supported in part by the Swiss National Science Foundation (SNF) under contract number
200020-175595 and by the Swiss National Supercomputing Centre (CSCS) under project ID UZH10.
This project received funding from the European Union's Horizon 2020 research and innovation programme
\textit{High precision multi-jet dynamics at the LHC} (ERC Condsolidator grant agreement No 772009).
RM was supported by STFC grants ST/S505365/1 and ST/P001246/1.

\bibliographystyle{elsarticle-num} 
\bibliography{ggdiphoton}

\end{document}